# Beyond flat-panel displays, applications of stereographic and holographic devices in 3D microscopy data analysis

Yong Wan, University of Utah, Salt Lake City, UT, 84112, USA

Holly A. Holman, University of Utah, Salt Lake City, UT, 84112, USA

Charles Hansen, University of Utah, Salt Lake City, UT, 84112, USA

*Abstract—Laser scanning microscopy enables the acquisition of 3D data in biomedical research. A fundamental challenge in visualizing 3D data is that common flat-panel displays, being 2D in nature, cannot faithfully reproduce light fields. Recent years have witnessed the development of various 3D display technologies. These technologies generally fall into two categories, stereography and holography, depending on the number of perspectives they can simultaneously present. We have integrated support for many commercially available 3D-capable displays into FluoRender, a visualization and analysis system for fluorescence microscopy data. This study investigates the opportunities and challenges of applying various 3D display devices in biological research, focusing on their practical use and potential for broad adoption. We found that 3D display devices, including the HoloLens and the Looking Glass, each have their merits and shortcomings. We predict that the convergence of stereographic and holographic technologies will create powerful tools for visualization and analysis in biological applications.*

Microscopy techniques capable of generating 3D scans across multiple channels and time points have become standard in biological research. Current visualization and analysis methods rely on flat-panel displays, such as LCD (liquid crystal display) and OLED (organic light emitting diode) screens. However, projecting 3D data onto a 2D image plane presents a fundamental challenge for data visualization and analysis, as it diminishes depth and structural perception. The ability to visualize 3D structures in a life-like manner can significantly enhance data comprehension, bypassing many software techniques developed to address the limitations of flat-panel displays. Given that the adoption of 3D display technology is largely driven by consumer demand for entertainment, we present a series of commercially available stereographic and holographic display devices and examine their potential for analyzing microscopy data in biological research. We integrated support of these display devices into a data analysis system dedicated to biological research, FluoRender. This system enables a fair comparison of various display devices in real-world applications. Our focus is on the practical use and potential for broad adoption of each device in biological research. Additionally, we identified several previously overlooked features, such as depth perception accuracy, resolution, and ease of use, which may present both opportunities and challenges for the widespread adoption of these technologies in the future.

# BACKGROUND

## 3D data acquisition by scanning microscopy

Laser scanning microscopy, including confocal and two-photon microscopy, is a powerful technique for acquiring 3D data from fluorescently stained samples. In confocal microscopy, a focused laser beam scans point-by-point across a plane in the sample to excite fluorescent dyes. The emitted light is collected through a pinhole aperture to eliminate out-of-focus light, resulting in high-resolution optical sections. By capturing a series of these sections at different depths (z-stacks), a complete volumetric data set is generated [1]. On the other hand, two-photon microscopy uses two photons of lower energy to excite the fluorescent dyes simultaneously. The process requires high-intensity pulsed lasers and allows for deeper tissue penetration with reduced photodamage and photobleaching. Unlike confocal microscopy, two-photon microscopy does not require a pinhole for optical sectioning, achieving higher clarity and larger volumes for thick tissues [2].

## Light field reproduction techniques

The light field is the portion of the full electromagnetic field that can be perceived by human eyes. The human brain processes visual information based on neural signals detected by the retinas of both eyes. Light field reproduction aims to create sufficient stimuli for the retinas to trick the brain into perceiving a three-dimensional scene.

**Intensity reproduction** refers to the process of replicating the magnitude of the electric or magnetic field at an image plane. This technique involves reproducing the light intensity of specific wavelengths that correspond to the photoreceptor cells in human eyes, on a plane or surface, such as photos and flat-panel displays.

**Structure from motion** involves basic user interactions such as rotating, panning and zooming, which are essential for flat-panel displays but also work with various 3D displays [3]. Additionally, head tracking, eye tracking, and hand tracking provided by various head-mounted displays utilize structure from motion to enhance 3D perception.

**Stereoscopy** leverages human binocular vision to present slightly different images to each eye, creating a sense of depth through parallax. The techniques used to create stereoscopic images are known as stereography, which has a history as old as photography.

**Directional distribution of light field.** A flat panel only reproduces the light intensity at an image plane. Each pixel acts as a point light source, generating spherical waves from a common plane. The interference patten and subsequent light propagation differ significantly from a real scene, where light sources are at different depths. Traditional holography reproduces both the light phase and intensity at an image plane, using a reference light and its interference pattern for encoding and decoding a complete light distribution [4]. This approach better mimics how light propagates off the image plane as in a real 3D scene.

## Commercial 3D displays

The following is a summary of the mechanisms behind current commercial 3D display technologies.

**3D TVs.** They utilize intensity reproduction and stereoscopy. Active or passive shutter glasses are needed to create the 3D effect.

**Nintendo 3DS.** They employ intensity reproduction and stereoscopy through a lenticular lens array, allowing users to experience 3D without glasses. We did not test them in this study because they have been discontinued and cannot function as standalone displays.

**Cardboard VR.** They use intensity reproduction and stereoscopy. Lenses are used to refocus images, providing an affordable and accessible way to experience virtual reality (VR).

**VR headsets.** Combining intensity reproduction, stereoscopy, and motion tracking, they incorporate more sophisticated lens designs than the Cardboard VR for overall compactness and enhanced user experience.

**AR headsets.** Augmented reality (AR) generally utilizes similar technologies in VR headsets but incorporates different waveguide designs to redirect light from virtually generated images while maintaining an open view.

**The Looking Glass.** They are similar to the Nintendo 3DS but use lenticular lens arrays for light modulation to achieve 3D object presentation at multiple views, effectively creating holography. This technology reproduces light directional distribution without the need for glasses, allowing for an intuitive 3D experience. It leverages stereoscopy, structure from motion, and light-field reproduction simultaneously.

# IMPLEMENTATIONS

We integrated support for a broad range of 3D display devices into FluoRender, a tool for visualizing and analyzing 3D, multichannel, and time-dependent fluorescence microscopy data in biomedical research [5]. Implementing various 3D display technologies within one platform allows us to compare them on a common ground.

## Cardboard VR and 3D TVs

Stereography is achieved by rendering each frame twice, once for the left eye and once for the right eye. The camera position is shifted for each eye, and the shifted distance can be set in the user interface. The left and right images are placed side by side in one image for full-screen display. The most common choices for the Cardboard VR display are Android and iOS phones with screen sizes ranging from 4 to 6 inches. These phones can be connected to the host computer via USB cable or wirelessly. Since most mobile phones' USB ports lack the support for video input protocols, a third-party program is usually needed to transmit video signals from the computer to the phone. Small form factor displays commonly used for Raspberry Pi devices can also be connected to the computer via USB or HDMI.

This implementation can be used with 3D TVs almost without modification. Most 3D TVs allow viewers to select a 3D frame format, with the side-by-side format being a convenient choice to stay compatible with the Cardboard VR. Since the side-by-side format compresses the left and right frames horizontally to fit both frames into one standard video frame, we adjusted the aspect ratio of the rendered frames to compensate for the compression.

## VR headsets

Major brands of VR headsets include Meta Quest, HTC Vive, Valve Index, and Windows Mixed Reality [6]. Each brand provides its own proprietary application programming interfaces (APIs) for communication between the headset and a computer. OpenXR, a cross-platform API provided by the Khronos group, is primarily used to support a wide range of VR headsets. Additionally, we employed the OpenVR API from Steam to improve cross-platform compatibility, as SteamVR supports various types of headsets and provides a reliable runtime for macOS. One caveat with using OpenXR is that it is independent of a specific graphics API binding, and a specific headset may lack support for the graphics API used by the host program. For example, Windows Mixed Reality headsets only support Direct3D as the graphics API. To successfully use these headsets with cross-platform APIs like OpenGL and Vulkan, we added an extra step in FluoRender to transfer the rendering results to Direct3D via shared textures.

There are two types of VR headsets: tethered and standalone. Tethered headsets are wired to a host computer to receive graphics content, while standalone headsets function as a mobile computer and can communicate with a host computer using a wired or wireless network. Windows Mixed Reality, HTC Vive, and Valve Index are examples of tethered headsets. Meta Quest headsets are standalone but provide the Quest Link interface to handle communications with a host computer without extra programming. Therefore, OpenXR can work with all these VR headsets.

We support render-view interactions via original FluoRender controls, such as using the mouse and keyboard to rotate, pan, and zoom the view, as the VR contents are mirrored in the FluoRender render view. Additionally, a game controller is supported using XInput under Windows. We also support built-in tracking and controls of VR headsets and accompanying controllers, implemented using OpenXR and OpenVR. For example, squeezing the grab button on a controller enables the mapping of the controller's pose to the 3D data, allowing users to rotate the controller and the data synchronously to examine the data from different directions.

## Microsoft HoloLens 2

HoloLens is currently the most adopted AR headset, with applications in industrial and military fields [7]. For AR program development, it functions as a standalone device, similar to Meta Quest, but is based on the Windows Mixed Reality API. Since Microsoft does not provide a transparent interface like Quest Link, its connection and communication with a host computer are achieved through either the proprietary API or an extension to OpenXR. We adopted the OpenXR extension in FluoRender so that all VR and AR functionality could derive from a base class using OpenXR, allowing us to add features as needed, such as Direct3D support and networking with HoloLens. HoloLens does not come with dedicated controllers; user interactions are achieved via hand gestures and voice commands, in addition to the original FluoRender controls using a mouse and keyboard.

We observed a limited revival of interest in AR headsets recently with the release of Apple Vision Pro and experimental products like Meta Orion. We believe many of the latest AR headsets bypassed some of the issues associated with HoloLens, offering more processing power, increased field of view, and better image quality. However, we could not test all of them, and we believe HoloLens still provides a representative experience of AR in biological applications.

## The Looking Glass

The process to generate renderings for the Looking Glass is comparable to lenticular printing, which uses a parallel convex lens array for easy manufacturing [8]. Images rendered from continuously changing viewing angles of a 3D object are divided into slivers of single-pixel width and reassembled to align with the lens array on the display (Figure 1). This process is embedded into the VR/AR rendering pipeline by increasing the number of images from two eyes to multiple views, generating a texture array of multi-view images. The image reassembly is achieved by lookups of the texture array with parameters describing the layout of the lens array provided by the display itself. Some technical details of the implementation are worth noting. 1) The lens array layout can vary from one display to another, with parameters determined by a calibration process during manufacturing. 2) The total number of images to render is flexible; numbers from 30 to 100 can work but influence the smoothness of transition from one view to the next. 3) The viewing angle of a portrait display is about 45 degrees horizontally. 4) The camera motion to generate these images of different viewing angles is turntable instead of shifting. Therefore, we typically render 45 images of 1 degree of rotation for each. The Looking Glass is connected to a computer as a normal display device. Therefore, we detach the render view from the standard FluoRender window and place it on the Looking Glass in full-screen mode. This allows us to retain all the original FluoRender interactions. Users can rotate, pan, zoom, and play a time-dependent sequence in the Looking Glass render view the same as standard FluoRender operations.

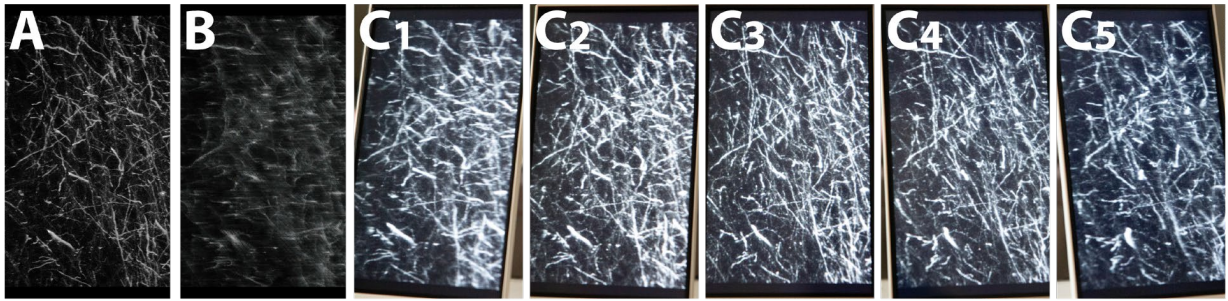

FIGURE 1. A scan of neuron fibers displayed on a Looking Glass display. (A) A screen capture without holography, where the orientation of a fiber is unclear due to the loss of depth information. (B) The assembled image containing renderings from 45 viewing angles. The result resembles a rotational motion blur. (C1-5) Photos of the display from five real-world viewing angles. The spatial orientations and relationships of the fibers are easily understood from the changes in perspective. Note that the photos of the Looking Glass display appear blurrier than what human eyes perceive because a camera lens has a larger light-receiving area, causing images of adjacent viewing angles to blend.

## Deployment

All functions for stereographic and holographic displays have been available for public download since FluoRender 2.32. Executables for different operating systems as well as source code can be downloaded from GitHub (https://github.com/SCIInstitute/fluorender). All display-related settings are under the Display tab of the configuration dialog in FluoRender (Figure 2). It allows users to switch on one type of display at a time.

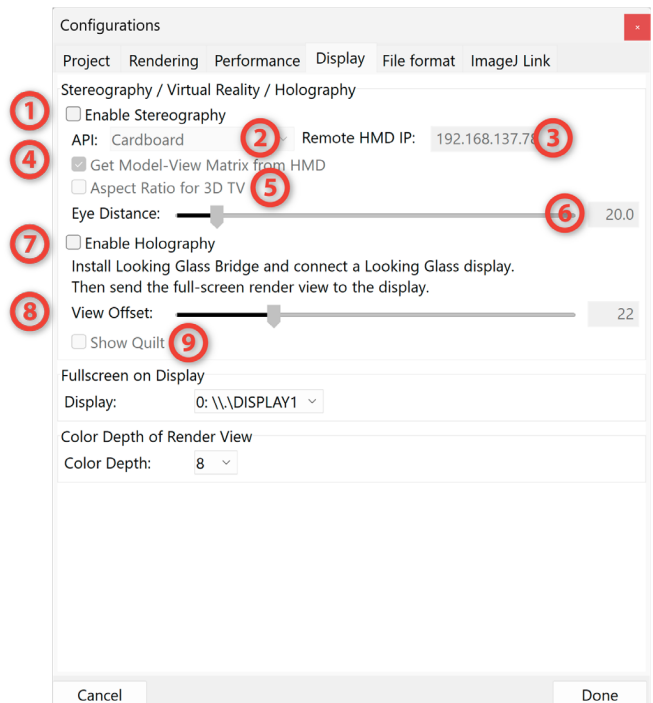

FIGURE 2. The user interface for setting up various 3D display devices in FluoRender. (1) Option to enable stereography. (2) Options to select an API for a connected device. (3) If a HoloLens is connected, set the IP address of the HoloLens. (4) Option to enable head tracking. (5) Adjust the aspect ratio for 3D TVs. (6) Set the eye distance in the virtual environment. (7) Option to enable holography. (8) Adjust the rotation angle for generating multi-view images. (9) Option to show the image array instead of the assembled image.

For a stereographic display, users need to connect and configure a supported device before enabling the feature in FluoRender. For VR headsets, this means that a VR runtime needs to be installed, such as SteamVR, Quest Link, or Windows Mixed Reality Portal. When no proper VR runtime is available, FluoRender will default to Cardboard mode, which simply renders left and right eye images side by side. The Cardboard mode can be used for a 3D TV, but an option to change the aspect ratio needs to be enabled. The OpenXR runtime for HoloLens is included within the FluoRender executable. Users need to install the Holographic Remote player on the HoloLens and connect to the same network as the host computer. Then, the IP address of the HoloLens is needed in the settings to allow network communication between the HoloLens and the host computer. For the Looking Glass, users need to install the runtime provided by the manufacturer, called the Bridge. Once a stereographic or holographic device is connected and configured correctly, switching on the corresponding option in FluoRender settings will automatically redirect renderings to the desired device.

We added support for these 3D display technologies over a long development time span, mostly in response to user requests and with consideration of device availability. We performed informal user studies through presentations, demonstrations, and direct communications. The development team of FluoRender worked with biologists and gained firsthand experience of how the 3D display technologies might be used in research in a practical manner.

## OPPORTUNITIES AND CHALLENGES

FluoRender is freely available to the public and has a substantial user base. However, the requirement for additional hardware to support 3D display technologies limits adoption of these devices. Conducting a large-scale survey to gauge user opinions on these 3D display devices is currently not feasible. Instead, we evaluate and compare these devices based on the authors' own use experiences and observations of biologists' responses during interactions with FluoRender users.

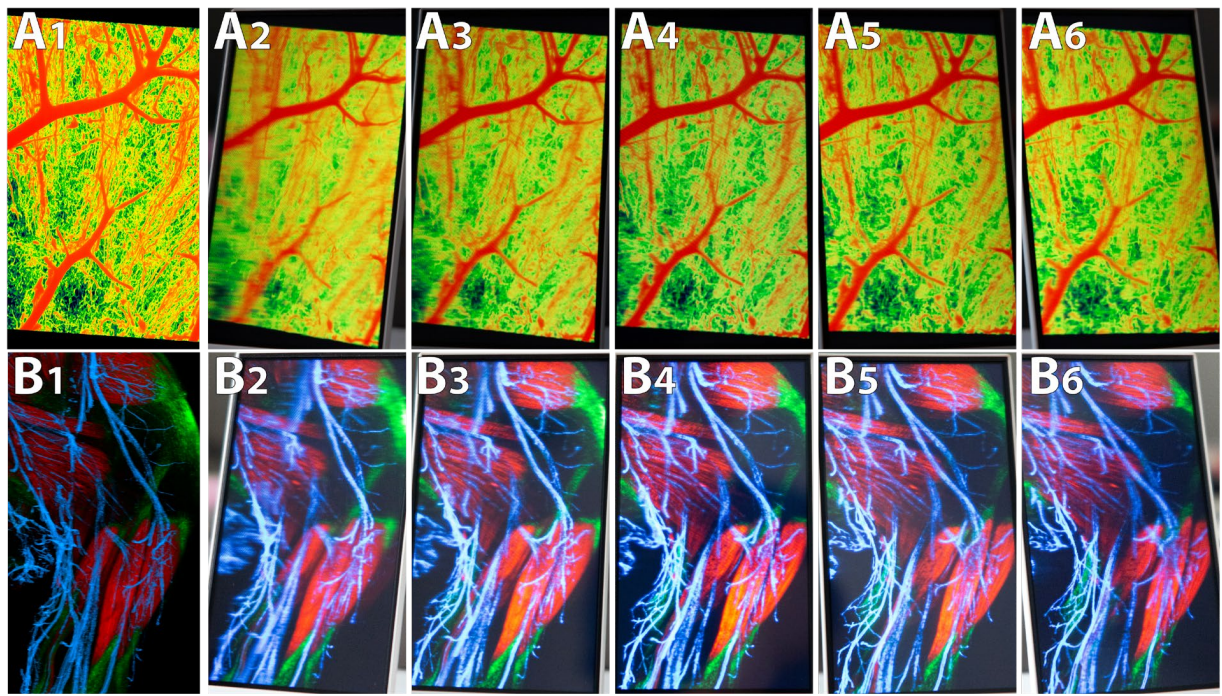

FIGURE 3. Depth perception enhanced by 3D display devices, demonstrated by photos of the Looking Glass. (A1) A mouse kidney scan rendered using MIP with intensity values mapped to colors. It is unclear which branching and fibrous structures are in front. (A2-6) On the Looking Glass, the spatial relationships are clear even for MIP visualizations. (B1) A mouse hindlimb scan of three channels rendered with channel layering; the nerves in blue are in the top-most channel. It is unclear how the muscles in red are innervated. (B2-6) The Looking Glass enables effortless understanding of muscle innervations as the relative locations of the structures shift with viewing angles.

## Enhancing depth perception

Before determining which device can best present depth information of 3D microscopy data to the user, we must first consider what an ideal visualization would look like for biologists. Ideally, biologists would expect a physical enlargement of the sample they want to examine, allowing them to freely and easily handle it to view the details. In this sense, AR devices, such as the HoloLens, aim to achieve this goal by leveraging stereoscopic vision, mixing it with the real environment, and tracking the motions of the head, eyes, and hands. Various VR devices can achieve similar results by incorporating both stereoscopic vision and structure from motion (usually head tracking and sometimes hand tracking) to present depth information. The holographic display, represented by the Looking Glass, is also effective because it utilizes stereoscopic vision and structure from motion more intuitively, although the range of viewing is limited to approximately 45 degrees horizontally. Stereoscopic displays without head tracking can only achieve limited realism, as they do not utilize structure from motion. For example, viewing 3D data on a 3D TV generally allows discrimination of depth order of different structures. However, when the viewer attempts to examine occluded structures by moving their head, the resulting perspectives are often interpreted as distortions of the data, as only two angles for the left and right eyes can be provided.

A noteworthy finding in applying stereographic and holographic displays to 3D microscopy data visualization is that techniques previously regarded as poor or even misleading for visualizing spatial relationships become especially useful when coupled with 3D display devices. **Orthographic projection** is commonly used for microscopy data visualization because it preserves the relative size of front and back structures. When a scale bar is displayed, it allows for quick measurement of lengths, which can be crucial in research. Orthographic projection is usually not used with various VR and AR headsets because the projection is determined by the optics in the headset. For non-head-mounted displays, notably the Looking Glass, users can enable orthographic or perspective projections. The ability to simultaneously present multiple viewing angles on the Looking Glass replaces the need to rotate the view with a mouse and allows for a quick understanding of spatial relationships through natural stereoscopic vision and head movements. **Maximum intensity projection (MIP)** is valuable for fluorescence microscopy data visualization as it emphasizes high-intensity structures, mostly the target of fluorescent labeling. Visualizing 3D data using MIP can be misleading because a high-intensity structure behind a low-intensity structure is shown at the front. Efforts to mitigate the shortcomings of MIP using various rendering techniques, such as adding shading and shadows, have been made [9]. However, with a display device capable of showing 3D data, users can now visualize using MIP without worrying about the loss of depth information (Figure 3A). In FluoRender, MIP can be enabled or disabled for all the 3D display devices under discussion. **Channel layering** is one of various intermixing methods for multichannel microscopy data, rendering each channel separately and layering them with a transparent background [10]. Channel layering is a straightforward method to emphasize one or specific channels placed on top. However, placing a channel in front of others and disregarding the inter-channel spatial relationships can lead to misleading visualizations, similar to MIP. This issue is resolved when a 3D display is used, as the depth order of structures from different channels is understood through parallax (Figure 3B).

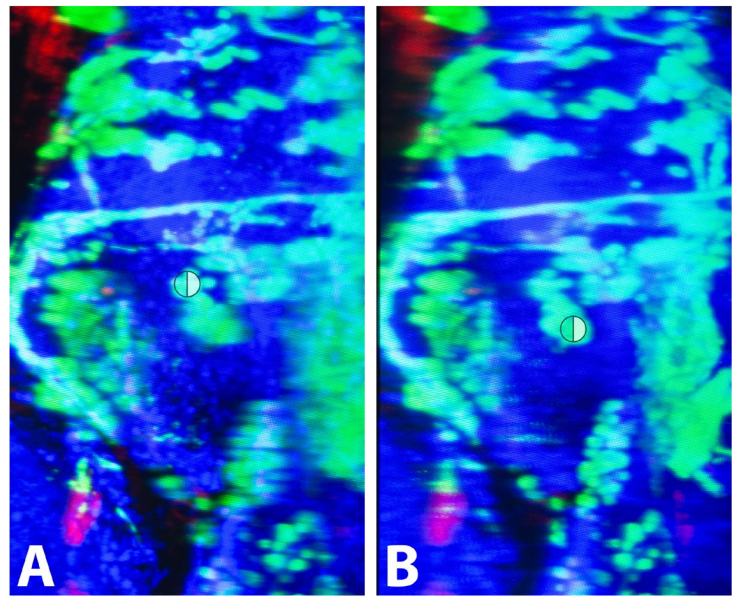

FIGURE 4. Two photos of the Looking Glass display, visualizing a zebrafish head scan. (A) and (B) are rendered using different rotation centers, computed from focus points indicated by the circles. The depth-of-field effect is evident as the clarity of structures at different depths shifts with the focus point.

Rendering with depth of field is a visualization technique that enhances depth perception and emphasizes important structures [11]. When multiple images from rendering a 3D data set by consecutively rotating viewing angles are presented on a Looking Glass display, images from adjacent viewing angles can be seen simultaneously and blend together. This effect is most noticeable for structures farther away from the rotation center. The results resemble the depth of field effect, where structures at the rotation center are sharp in focus, and the visualization transitions into blur as structures extend farther from the focus point. To leverage this phenomenon, we implemented an auto-focusing function. We set the center of the render view as the focus detection point and use ray casting to compute the distance to the first-hit structure [12]. This closest structure is then used as the rotation center for generating the turntable renderings for the Looking Glass. This approach facilitates viewing details of a large microscopy scan, as the center of the view remains clear when users pan and rotate the data (Figure 4).

## Frame rate and motion sickness

When implementing rendering pipelines for various VR headsets, it is recommended to maintain a frame rate of at least 90 fps to avoid motion sickness. A common strategy is to reduce detail by downsampling the data or reducing the output frame buffer size when the desired frame rate cannot be achieved for rendering a large volume. However, this can result in an degraded and less useful visualization. Conversely, if we choose to maintain the visual quality and allow the frame rate to drop significantly below the recommended target, motion sickness does not seem to impact users, based on our observations. We collaborated with researchers specializing in the vestibular system to investigate this matter further. The primary cause of motion sickness from VR is vestibulo-ocular reflex (VOR) induced by the mismatch of the visually perceived motion from that detected by the vestibular system in the inner ears [13]. For example, the most severe motion sickness is induced by viewing a VR video of a moving camera or walking around in a VR scene while stationary. Motion sickness is unavoidable by current VR technologies regardless of frame rate due to the significant discrepancy between the motion detected by the visual and vestibular systems. This is why most VR games confine user actions within a static, room-sized environment, where user movements are synchronized between the virtual and real environments; teleportation is often employed to cover longer distances to transition from one static environment to another. For viewing biological data, there is generally no such environment as a visual reference for movements. We commonly use a solid color for the render-view background, which is intuitively interpreted as a static environment by the user. Therefore, even if the frame rate of VR rendering drops and motion stuttering becomes obvious, the user experience does not differ much from viewing the visualizations directly on a flat-panel display. Without the need to present a virtual environment, VR headsets are better suited for biological data visualization than gaming. For the same reason, AR headsets adopt see-through lenses to use the real environment and avoid visual-vestibular mismatch altogether.

High frame rate is not required for AR headsets. For example, the maximum refresh rate of the HoloLens is 60 Hz.

Rendering dynamic data on the Looking Glass while maintaining an interactive frame rate is a significant challenge. Typically, we render 45 images from different viewing directions to assemble one hologram for the Looking Glass. For viewing a static data set, the Looking Glass offers an intuitive and convenient experience. The image sequence only needs to be updated once, allowing users to observe 3D structures through stereoscopic vision and head movements. However, when a user wants to rotate the data set to view its back or visualize a time sequence, the update becomes time-consuming due to the brute-force rendering method. Currently, we employ a streamed pipeline and update the result as each image for each viewing angle finishes rendering. This results in a gradual update of the 3D visualization during viewport manipulations and time-sequence playbacks. It would be interesting to explore novel rendering algorithms to accelerate the interaction speed for the Looking Glass. We expect this can be approached by first reducing the size of the volume data by exploiting redundancy at data level and then mitigating the efforts to render images from different angles by leveraging the recurring patterns in the images.

## Resolution reexamined

Current flat-panel displays have reached the sweet spot of human vision resolution at normal viewing distances due to their sufficient pixel density, which is true for various panels, including TVs, phones, and tablets. Although the devices we examined have diverse form factors and builds, almost all of them are built on top of conventional flat-panel displays. Therefore, their raw resolutions should be sufficient for visualizing microscopy data with clarity. However, the techniques that make normal display panels 3D capable also sacrifice perceived resolution. More sophisticated 3D display solutions usually incur a greater penalty in image clarity. We first examine a 3D TV with FluoRender working in the Cardboard VR mode. The resolution is halved (50%) when two images are compressed to fit within one frame. The perceived resolution loss is smaller because the two images are highly correlated. An empirical estimation suggests about 70% of the information is retained for the 3D TV, which is similar to interlaced frames used in video encoding.

The rendering for Cardboard VR does not reduce the display's pixel density per se. Some phones use 4K display panels, and using them in Cardboard mode can achieve the best perceived resolution in our experience. However, images are still pixelated because the viewing distance is extremely short. The serious impact on the resolution of VR headsets, in general, is the low utilization of human foveal vision. For example, each display on a Meta Quest 2 has a resolution of 1832 x 1920 pixels and covers a 97 x 93 degrees field of view for each eye. Considering the coverage of the foveal region of human retina is only about 2 degrees, the pixels that can contribute to human foveal vision are about 38 x 41 [14]. As discussed previously, VR headsets are designed to present a full environment, as in VR games and videos. Biological data visualization demands a focused examination of structural details instead of a full environment. When a data set is visualized at a distance to utilize the foveal vision, the results are too pixelated to show its details. This forces users to zoom in on the data, which in turn covers a large field of view and requires the user to turn their head to look at different parts. The process can become counterintuitive as it differs from how a data set is normally examined. An AR headset like the HoloLens concentrates its display pixels in a narrower field of view because rendering the full environment is not needed. The HoloLens has 1440 x 936 pixels covering 43 x 29 degrees for each eye, resulting in 67 x 66 pixels covering the foveal vision. The HoloLens is the only device not based on flat panel display technology. It uses a Micro-Electro-Mechanical Systems (MEMS) scanner to redirect laser into the retina for each eye. The use of mechanical parts also limits the bandwidth of the display, making the overall resolution much lower than a typical VR headset. However, the perceived resolution of the HoloLens is clearly higher.

The Looking Glass includes a product line of various display sizes and resolutions. The most affordable models are based on large phone-sized display panels of 6 or 7 inches with resolutions ranging from 2K to 2.5K. Their high-end models increase display size and resolution proportionally. With a typical viewing distance, the perceived resolution of 3D data is higher than the VR and AR headsets we tested because the screen pixels are concentrated to cover the foveal vision. The Looking Glass simultaneously presents multiple highly correlated images. There is some resolution reduction from the multi-image assembly process, and we estimate that at least 30% of the information is retained. Considering the compression of the images mainly occurs horizontally, an estimated pixel contribution to the foveal vision of the Looking Glass Go (a portable model of the Looking Glass) when viewed at an arm's length is 126 x 412 pixels. The perceivably higher resolution of the Looking Glass displays for viewing 3D microscopy data aligns with the user experience from several informal user studies.

### Ease of use

All 3D display devices require specialized hardware connected to a host computer, and many also need additional software to function properly. Their adoption in biological applications is significantly influenced by the convenience of their utilization. Although the measure of convenience is subjective, we can draw some general conclusions by comparing them in typical application scenarios. First, we establish a baseline for how an average user employs FluoRender for typical microscopy data analysis: using a flat-panel display connected to a desktop computer. Then, we examine the additional effort required to use 3D display devices.

A 3D TV can seamlessly replace a standard flat-panel display in the existing setup. To view 3D data on a 3D TV, a user needs to enable settings in FluoRender, adjust the TV settings (side-by-side stereography), and wear 3D glasses. Active shutter glasses require more operations than passive polarizing glasses. Since 3D TVs are usually larger than typical desktop displays, a comfortable viewing distance is longer than in a standard work environment. Room space and remote operation of the host computer should also be considered.

A phone-based Cardboard VR setup requires similar efforts as a 3D TV. The weight and fit of the phone housing can make users feel claustrophobic, leading to discomfort. A proper VR headset generally requires significantly more effort to set up. They also need a clutter-free workspace to allow free body movements and avoid collision with obstacles. Wirelessly connected and inside-out tracking headsets like the Meta Quest series can ease some efforts, such as cable management. However, using VR headsets for routine data analysis is nearly impractical as it requires frequent removal to accomplish tasks in the real environment. The delays from connecting and reconnecting the VR headset can make the workflow cumbersome.

An AR headset like the HoloLens addresses many limitations of VR headsets that may deter users from using them at workplaces. An AR headset is generally lightweight and keeps an open view of the environment. While most users agree that the HoloLens is more comfortable and convenient than the Meta Quest headset, they also notice the shortcomings of AR headsets in practice. First, it is recommended to have an environment free from visual clutter, ideally dark-lit with a blank wall, for viewing the 3D data details on an AR headset due to the overlapping of virtual images and real environments. Second, the lenses of the HoloLens are waveguides for redirecting laser beams to the eyes, which can distort the perceived real-world environment, especially the contents on a normal LCD display. Constantly wearing the HoloLens does not guarantee fluid workflows that require frequent switching between AR visualizations and routine tasks on a flat-panel display.

3D display technologies like 3D TVs, Cardboard VR, VR headsets, and AR headsets all require wearable devices, making them personal. Ideally, these devices are not suited for sharing among multiple users due to hygienic concerns. Additionally, some devices need to be set up to match the user's interpupillary distance (IPD) and may require spacers for prescription glasses. The HoloLens, for instance, scans the user's iris for security and requires eye calibration to display stereography correctly. These factors make the AR and VR devices inconvenient in a collaborative work environment. Furthermore, they require extra care post-utilization for cleaning, recharging, maintenance, and storage.

A holographic display like the Looking Glass is superior in terms of its seamless integration into existing data analysis workflows. The Looking Glass can be constantly connected to the host computer as a secondary display, posing no obstruction to routine work. With FluoRender, 3D data are automatically visualized and examined on the Looking Glass when the option is enabled. Many users have found this process highly effective, allowing them to quickly switch to 3D visualization and uncover spatial relationships between structures in microscopy data without explicit viewport manipulations. A potentially influential factor we were unable to examine is the fatigue caused by extensive use, which would require long-term tracking of user experience. We anticipate that all 3D displays will cause user fatigue to varying degrees.

We found that user demand for the ease of use extends beyond direct utilization of these devices. While users agree that most 3D displays can enhance their perception of depth information and facilitate new discoveries, they often ask: how can we use these devices for publications? The fact that none of the 3D display technologies can be easily integrated into common formats for sharing research findings, such as printed media, electronic documents, and videos, significantly dampens enthusiasm for their adoption in practice. Sharing scientific findings visualized using 3D display technologies generally requires the audience to be equipped with the same devices as the authors of the publication. For the Looking Glass display, we hypothesized that traditional lenticular printing might be incorporated into the publication process. For example, a standard lenticular sheet can be distributed to subscribers of a journal that publishes encoded images in printed or digital formats. Users could then overlay the lenticular sheet as an optical decoder to view the 3D data. However, quality assurance becomes impractical, as even the slightest change in magnification and orientation of the images can result in misalignment and invalid visualizations. Like this article, we can mostly describe our experience with each 3D display device instead of showing the results directly. Even for the Looking Glass, which does not require wearing glasses, capturing its clarity through photos remains challenging. We invite readers to experience these devices firsthand to appreciate their visual impact. Furthermore, we encourage FluoRender users to share their data directly. Other users who manage to obtain any of the supported 3D display devices can then view the shared data on a device of their choice.

## Cost, availability, and longevity

We will organize our discussion based on the current market prices of the 3D display devices.

**Cardboard VR.** Assuming a user already owns a phone, which is excluded from the cost calculation, Cardboard VR housings are produced in large quantities with inexpensive retail prices. They can be easily found at most electronics retailers. Users only need to ensure the housing size fits their phone models. Their long-term availability is linked to mobile phone formats, which are expected to remain relatively stable for the foreseeable future.

**VR game headsets.** Prices range from $300 to $1500. VR headsets have seen rapid growth in recent years, but the market for VR gaming remains specialized and niche. Contemporary models from major manufacturers are easily available. However, some previously popular models have been discontinued, such as Microsoft's Windows Mixed Reality headsets, which are no longer supported in recent Windows updates. The limitations of VR headsets have hindered their mass acceptance, leading some manufacturers to shift focus to AR headsets for daily and personal use.

**3D TVs.** Generally priced similarly to non-3D flat-panel TVs, 3D TVs have become less common in recent years. Major manufacturers have largely discontinued production due to waning consumer interest and limited 3D contents, such as 3D Blu-ray movies. A revival of the 3D TVs seems unlikely in the foreseeable future.

**Small to medium form factor Looking Glass displays.** Prices range from $300 to $4000. All models can be purchased directly from the manufacturer. Lenticular-based holographic displays are exclusive to the Looking Glass Factory, which markets them as 3D photo frames or for product demonstrations. Although we think they are well-suited for integration with existing data analysis workflows in biological research, holographic displays will likely remain a niche market for some time.

**AR headsets.** Major products currently include the Microsoft HoloLens 2 and Apple Vision Pro, with retail prices around $3500. Both products are being phased out, and the future of their respective product lines is uncertain, making it increasingly difficult to procure these devices.

**Large form factor Looking Glass displays.** Prices and availability can be obtained from the manufacturer upon request. Currently, the costs are prohibitive for practical use or broad adoption.

## FUTURE PREDICTIONS

We expect that the two major camps in 3D display technology – stereography, as seen in various VR and AR headsets, and holography, represented by the Looking Glass – will continue to develop. Head-mounted displays will remain the main form for stereography. Since the image-producing device can be placed relatively close to the eyes, the need to reproduce light directional distribution is less crucial for 3D scene reconstruction. Technological iterations in AR glasses development will enable further component miniaturization and optimized optical paths through innovative waveguide designs, leading to devices with higher resolution, lighter weight, and distortion-free see-through optics. Manufacturers are focusing on AR glasses development with the goal of creating popular personal devices in the form of ordinary eyewear. If this trend proves successful, many people will carry AR glasses with them, just like they carry phones today. This means scientists will have 3D-capable display devices readily available at any moment, greatly enhancing data visualization. On the other hand, optical waveguides are expected to remain the dominant method for practical holography. The Looking Glass can be improved in several ways. Incorporating a touch display design will enhance usability for interactive 3D data viewing. We anticipate an increase in resolution to compensate for information loss from the parallel lens array. Designs of lens organizations beyond just a parallel pattern will expand viewing directions beyond the horizontal.

Furthermore, liquid-crystal-based light phase modulation has the potential to resolve the resolution and viewing angle challenges of the Looking Glass [15]. If affordable light intensity and phase modulations can be integrated into display panels, a wide range of devices, including TVs, mobile phones, and head-mounted displays, could faithfully reproduce light field, effectively becoming holographic. This advancement is especially promising for AR glasses, as existing transparent-panel technologies can be adapted for holographic displays, eliminating the need for complex waveguide designs and enabling truly slim and lightweight eyewear. We foresee that the convergence of stereography and holography, driven by emerging technologies, will revolutionize display devices, transforming them into powerful tools for gaining deeper insight into 3D data in research. We predict a significant overhaul of graphics APIs to simultaneously handle light intensity and phase for holographic displays. The evolution of technologies is likely to follow a path of continuity, which is why the future graphics pipeline will resemble current graphics APIs, featuring a volumetric rasterization pipeline based on spectral methods [16], complemented by ray-tracing techniques for additional effects. Visualization-based data analysis will evolve in response to this fundamental shift in display and graphics technologies. Innovative techniques, such as the auto-focusing and channel layering methods we proposed in this article, will be required for data visualization to fully leverage depth information in holographic renderings. Additionally, we anticipate significant advances in interactive data analysis, with many data picking and filtering tools needing upgrades to facilitate more intuitive handling.

# ACKNOWLEDGMENTS

This work was supported in part by NIH R01 DC006685 and R01 EB031872. We wish to thank the individuals and organizations who provided the microscopy data for this study: Pavol Klacansky, 3Scan, Inc., A. Kelsey Lewis, Gabrielle Kardon, and Hideo Otsuna.